\documentstyle[aps,epsf,prd]{revtex}
\begin{document}
\begin{flushright}
IISc-CTS-02/03\\
hep-ph/0304061
\end{flushright}
\begin{center}
{\bf Review of chiral perturbation theory}
\footnote{Invited talk at the workshop, QCD2002,
Indian Institute of Technology, Kanpur, November 18-22, 2002}
\vskip 1cm
{B. Ananthanarayan}
\vskip 0.5cm
{Centre for Theoretical Studies, \\
Indian Institute of Science, Bangalore 560 012}
\end{center}
\begin{abstract}
A review of chiral perturbation theory and that of
recent developments on the comparison of its predictions with experiment
is presented.  Some interesting topics with scope for further
elaboration are touched upon.
\end{abstract}
\section{Introduction}
Quantum chromodynamics (QCD)
(for a review, see, e.g.~\cite{marpag}),
is the microscopic theory of the
strong interactions.  It is formulated for quark and gluon degrees of freedom,
and its fundamental coupling constant is the strong coupling
constant $\alpha_s$.  The theory is asymptotically free, in that as
a renormalizable field theory, the coupling constant becomes small at
large momenta, due to the non-abelian nature of the
gauge interactions and due to the fact that the number of quark flavors
is relatively small.  On the other hand, a Landau singularity at low energies
renders the theory non perturbative.  Furthermore, the asymptotic
spectrum of the theory consists of mesons and baryons, which implies that
a full solution to QCD must include the phenomenon of confinement of quarks
and gluons and ensure that these  emerge as the asymptotic
states of the theory.  When the energies under consideration are small 
enough for the $t, \, b, \,c$ quark degrees to have frozen out,
striking features of the spectrum include the fact
that some of the mesons are very light compared to the baryons.
One way to understand this, due to Nambu, is that of spontaneous breakdown
of approximate chiral symmetries of the strong interaction Lagrangian.
The members of the pseudo-scalar octet are the Goldstone bosons associated
with these spontaneously broken symmetries, and the $\eta'$ would be
the ninth Goldstone boson associated with the additional $U(1)_A$ symmetry
that would be realized when $N_c$ the number of colours tends to infinity.
In the real world where $N_c=3$, this symmetry is anomalous and 
therefore the $\eta'$ is very far from being a Goldstone boson. 

Chiral perturbation theory is the effective low-energy theory of
the strong interactions and is an expansion of the Green functions of
its currents associated with the near masslessness of the 3 lightest quarks, 
and with the spontaneously broken approximate axial vector symmetries,
in powers of momentum and quark masses.  The modern formulation of the
subject was presented in an influential series of 
papers for the case of 2 flavours in~\cite{classics1} and
for the case of 3 flavours in~\cite{classics2}.  
For a pedagogical introduction,
see~\cite{ss1}, and for detailed reviews, see~\cite{reviews}.
It is also possible to include
the electromagnetic and weak interactions into the framework, in order to
describe the effects of virtual photons and in order to study semi-leptonic
decays of mesons.  The nucleons and the baryon octet may also be included
in the theory that results in baryon chiral perturbation theory.  The
relativistic formulation was first presented in ref.~\cite{gss}.  Other
versions include heavy baryon chiral perturbation theory~\cite{jm},
and the recent infra-red regularized baryon chiral perturbation 
theory~\cite{bl}.

In the mesonic sector, at leading order one starts out with the non-linear 
sigma model,  and employing the  external field technique, one arrives at the
next to leading order in  the momentum expansion with an effective 
Lagrangian whose coupling constants absorb the divergences that are generated
by the non-linear sigma model Lagrangian at one-loop order.  This
procedure has also been carried out to the next to next to leading order,
and can in principle be carried out to arbitrary order.   At the formal level,
an important invariance theorem has been proved by Leutwyler~\cite{hl},
where it is shown that in order to arrive at a consistent formulation for
the generating functional, one must necessarily consider the symmetries
of QCD (the underlying level) at the local level, and establish the gauge
invariance of the generating functional for non-anomalous symmetries.
Note that the coupling
constants at each order have been to fixed by comparison with experiment.
Recently there has been some interest in trying to measure some of
these constants on the lattice, see, e.g.~\cite{hsw}.

For instance, at lowest order, the effective Lagrangian in the 2-flavor case
with mesons alone, which involves only the pion mass and the
pion decay constant, reads:
\begin{eqnarray*}
& \displaystyle {\cal L}_M={F^2\over 4} \langle
\partial_\mu U \partial^\mu U +M^2(U+U^\dagger)\rangle, &
\end{eqnarray*}
where $U\in SU(2)$ and $\langle A \rangle$ denotes the trace.  The matrix
$U$ contains the pion fields, and $F$ is the pion-decay constant.
The mass of the pion is such that $M^2=2 \hat{m} B$, where $\hat{m}$
is the average of the u- and d-quark masses, and $B$ is the value of
the quark condensate, the leading order parameter that determines the
spontaneous symmetry breaking of the chiral symmetry.  Today the accurate
measurements of pion scattering lengths, to be reviewed later, have confirmed
this to be the case, something that had been called into question~\cite{kms}.
At a purely theoretical level, if the quark condensate had indeed vanished,
one would have to find other order parameters which would have led to
sponaneous symmetry breaking.  These would have to higher dimension operators
such a s a mixed condensate of dimension 5 involving quark and gluon
field tensors, or those of higher dimensions, which are less and less
theoretically appealing, but distinctly allowed by the theory.

In the following sections, we will review some of the important and
interesting advances that have been made in the formalism, computation
and comparison of theory with experiment.

\section{Meson physics}
Chiral perturbation theory achieves its best results in the purely
mesonic sector.  Furthermore, when one confines oneself to the three
lights mesons, an unprecedented level of accuracy in the low-energy
sector can be reached, which provides a great challenge also to experiment.
As far as the formalism is concerned the complete enumeration of the
terms of the effective Lagrangian at two-loop or $O(p^6)$ order is
now available~\cite{bce1}.  The formidable task of renormalizing the
theory at this order is also complete~\cite{bce2}.    
This is achieved by starting out with the lowest order Lagrangian
and expanding around its classical solutions to generate the
loop-expansion of the generating functional.  The divergences of
the one-particle irreducible diagrams are always local and
can be renormalized.  The determination of the divergences turns out
be the evaluation of Seeley-DeWitt coefficients and the reduction of
the result to a standard basis of operators.  The result is checked
by the requirement of the Weinberg consistency conditions, which state
that the residues of the poles in $(d-4)$, in dimensional regularization,
should be polynomials in external momenta and masses.
The anomalous sector has also been considered, see~\cite{efs}.
The role of the scalar, vector, axial-vector, tensor resonances is
also well understood in the modern framework, and in general the bulk of
the contribution to the low-energy constants come from resonance saturation,
and has been elaborated in~\cite{egpr}.

\subsection{$\pi\pi$ scattering}
Pion-pion scattering is perhaps the simplest of all hadronic scattering
problems, since the pions are the lighest hadrons.  In the absence of 
electromagnetic interactions, the pions do not have a bound state, and
as a result, dispersion relations can be written in a simple and straightforward
manner for the system and are of great power for the analysis of
experimental information.  On the other hand, this setting is also the
one where chiral perturbation theory is expected to achieve maximum
accuracy.  This interplay of pion-pion
scattering amplitudes in chiral perturbation theory and their dispersion
relation representation has now been studied in great detail~\cite{cgl1}.

The starting point of this recent work is the computation
of the amplitudes to two-loop precision~\cite{bcegs,kmsf}.  The 
scattering amplitudes at this order involve two additional low-energy
constants over those that enter the expressions at one-loop order.
A systematic analysis of pion-pion scattering information 
available at medium and higher energies, far away from the threshold region
relies on a system of integral equations for the partial waves, the
Roy equations~\cite{Roy}.  Since dispersion relations for pion-pion scattering
amplitudes in the s-channel converge with as many as two-subtractions,
fixed-t dispersion relations bring in unknown t-dependent functions.
Using crossing symmetry effectively, these functions may be eliminated in
favour of the two S- wave scattering lengths $a^0_0$ and $a^2_0$
(note that there are three iso-spin amplitudes with $I=0,1,2$, generalized
Bose statistics imply that for even $I$, $l=0,2,4,...$, and for $I=1$, 
$l=1,3,5,...$; and the partial wave expansions for the
iso-spin amplitudes given by
\begin{eqnarray*}
& \displaystyle T^I(s,t,u)=32 \pi \,\sum (2 l +1) f^I_l (s)
P_l\left({t-u\over s-4 m_\pi^2}\right), &
\end{eqnarray*}
where $s, \, t, \, u$ are the conventional Mandelstam variables and
the $f^I_l(s)$ are the partial wave amplitudes,
such that we may introduce the threshold expansion in
the physical region $Re f^I_l(q^2)=q^{2l}(a^I_l + b^I_l q^2+...), 
q^2=(s-4 m_\pi^2)/4$).  At leading order, the predictions for the
two S- wave scaterring lengths is the one given by Weinberg from
current algebra and these read $a^0_0=7 m_\pi^2/32 \pi F_\pi^2
(\simeq 0.16)$ and
$a^2_0=-m_\pi^2/16 \pi F_\pi^2$ respectively.
Once the Roy representation for the amplitudes is so obtained, the
Roy equations for the partial waves are obtained by expanding the
amplitude and the absorptive part of the amplitude in terms of
partial waves.  Thus the Roy equations for each partial wave involves
the scattering lengths (for $l\leq 2$), and the imaginary parts alone of all
the partial waves in the physical region
and therefore allows a reconstruction of the entire
amplitude from knowing the scattering lengths and the imaginary parts
of the partial waves in the physical region.
We note that in the t-channel, the $I=1$ amplitude is such
that a convergent dispersion relation with one subtraction can be
written down, which renders the combination $2 a^0_0-5 a^2_0$ into
a tightly constrained quantity.
By assuming that in the low-energy region, the
scattering is dominated by the S- and P- waves, one may turn the Roy equations
for these into a closed system, absorbing the high energy and higher angular
momentum state contributions into inhomogeneous terms for these.
(It may also be noted that upto and including two-loops in chiral perturbation
theory, the amplitudes have cuts generated only by the S- and P-
waves.)  The medium energy data in the form of phase shifts comes from
a variety of sources including the reaction $\pi N\to \pi\pi N$,
the accurate measurement of the $\rho$-shape from the electromagnetic
form factor of the pion from the reaction
$e^+e^-\to\pi^+\pi^-$ performed by the CLEO collaboration.  Other sources could
be the decay $\tau\to \pi\pi \nu$.

In the modern context, a fresh Roy equation analysis with the view of
combining dispersion relations with chiral perturbation theory has
been carried out~\cite{acgl}.  
The evaluation of the inhomogeneous terms, the
so-called `driving terms' for the Roy equations
requires a detailed analysis of the D- and F- waves from threshold up to
the beginning of the asymptotic region, of all the waves in the asymptotic
region from Regge dynamics and from Pomeron contributions,
and the treatment of crossing constraints on absorptive parts of the amplitude,
which leads to their determination to rather small uncertainties.

This Roy equation analysis from which certain low-energy constants
of the chiral expansion
are accurately determined,  and together with computations in chiral 
perturbation theory both for the pion scattering amplitudes and for its
scalar form-factor yields a precise two-loop prediction for the 
scattering lengths has been carried out,
giving $a^0_0=0.220\pm 0.005$~\cite{cgl2}.   
One of the most significant experiments is one where the
phase shift difference
$\delta^0_0-\delta^1_1$ in the near threshold region is probed by the
rare decay $K_{l4}$.    The E-865 experiment at Brookhaven
National Laboratory has reconstructed 400, 000 events 
and reports a value $a^0_0=0.229\pm0.015$~\cite{e865}, 
which is to be compared with the 
results from the Geneva-Saclay experiment of 1977 based on 30, 000 
reconstructed events, which yielded $a^0_0=0.26\pm 0.05$.  

It may therefore be concluded that,
what is probably the problem studied in greatest detail in dispersion relation
theory, based on axiomatic field theory, is now at a stage where experiment
and theory are in full agreement after decades of analysis.  This remarkable
agreement also rules out a theoretical possibility that the leading order
parameter of spontaneous symmetry breaking is not the quark condensate
which would have implied a reordering of the chiral expansion, and a 
significantly larger value for $a^0_0$, perhaps close to even $0.30$.
In other words, the E-865 data leads to a precise measurement
of the quark condensate~\cite{cgl3}, and is therefore a sensitive probe
of the ground state of QCD.  

\subsection{Pionium lifetime computations}
Since it has always been a challenge to find experimental settings in which
pion scattering phenomena can be studied, due to the absence of pion targets,
it is important that the results from the $K_{l4}$ decays be independently
checked.  One system that affords such a test is that of pionium, where 
the decay of the atom made up of $\pi^+$ and $\pi^-$  bound
by the electromagnetic interaction, decays into a final state with 2$\pi^0$'s.
The lifetime of the atom gives a direct measurement of the combination of
scattering lengths: $a^0_0-a^2_0$.  This was first considered in~\cite{deser},
and is referred to as Deser's theorem.  In order to estimate the lifetime
at the desired level of precision, a bound state non-relativistic theory has
been formulated~\cite{glrg}, which has developed further a formalism first
proposed by Caswell and Lepage.  On the experimental side, it is expected
that the DIRAC experiment at CERN will reach the desired level of precision.

\subsection{$\pi K$ scattering}
The process in the full $SU(3)$ theory, which also entails an expansion in
the mass of the s-quark, that probes the structure of the theory is that
of pion-Kaon scattering.  At the formal level, the problem is more complicated
than that of the pion-pion case, since the process is inelastic due to the
$\pi\pi\to KK$ process in the t-channel.  The one-loop chiral amplitude
was first presented in~\cite{bkm}.

Indeed, as in the pion-pion scattering case, here too the amplitudes up to
two-loop accuracy have cuts generated only the the S- and P- waves, which
now number six.  By considering a system of amplitudes that are linear
combinations of the iso-spin amplitudes, that are even and odd under the
interchange of the Mandelstam variables $s$ and $u$ ,the dispersive
representations for the amplitudes have been written in a manner whereby
this can be compared with the chiral representations for the same amplitudes.
The corresponding dispersion
relation representation is technically more complicated, and the
economical Roy representation that was available for the pion-pion case
which exploited three-channel crossing symmetry is no longer available.
Instead, one turns to the Roy-Steiner representation for the 
dispersion relations
which employs both fixed-t and hyperbolic dispersion relations in order to
be able to perform a comparison with the chiral representation~\cite{ab}. 

On the experimental side, there is
significantly less precise data available.  For a review
of the experimental and theoretical scenario up until the
1970's, see~\cite{lang}.  While there are more recent
measurements of the S- and P- wave and some higher wave phase shifts for 
the $I=1/2,3/2$ amplitudes, in the t-channel the $I=0,1$ waves are also 
known in the physical region to varying degrees of accuracy.  In
the unphysical region $ 4 m_\pi^2 \leq t \leq 4 m_K^2$, the waves have
to be reconstructed from generalized unitarity and using Omn\`es
techniques.  While a full Roy equation analysis is in 
progress~\cite{bumo}, an analysis
based on exisiting information has led to an estimate for the
large $N_c$ suppressed low-energy constant $L_4$~\cite{abm}, which had been
estimated in the past only from Zweig rule arguments.  
A complete analysis combined with accurate experimental information
would also shed light on the important question of possible flavor
dependence of the quark condensate, and would be a sensitive probe
of the QCD ground state in the $SU(3)$ theory.
On the experimental
side it is hoped that pion-Kaon atoms can be studied which will lead to
a measurement of the scattering length $a^-_0(=(a^{{1\over2}}_0-
a^{{3\over 2}}_0)/3)$, and that the S- and P-
waves s-channel can be better measured at, e.g., the COMPASS experiment
at CERN.  The t-channel waves are discussed in some detail in ref.~\cite{mo}
whose considerations must be included in future analyses.

\section{Inclusion of weak and electromagnetic interactions}
From the earliest days of chiral perturbation theory the form factors for
various semi-leptonic decays have been considered, in order to study the
agreement between theory and experiment, for a review
see, e.g.~\cite{bceg}.  There has been considerable amount
of work for the measurements to be performed at DAFNE, and for the KLOE
detector.  

Recently at the formal level there has been interest in including
the effects of virtual photons, which would correct the masses for the 
pseudo-scalars and would also test well-known theorems established during
the days of current algebra calculations such as Dashen's theorem, 
which says that in the chiral limit, $m_{\pi^+}^2-m_{\pi^0}^2=
m_{K^+}^2-m_{K^0}^2$.
Introducing a chiral power counting in this sector leads to additional
terms in the effective lagrangian at $O(e^2 p^2)$.  The corresponding
effective lagrangian has been evaluated~\cite{urech}, and phenomenological
implications have been studied, e.g.,~\cite{nr,ku}.  There are now
additional low-energy constants associated with this effective lagrangian,
and there is very little knowledge on the magnitudes of these
quantities, and their determination
is a matter of current research, see, e.g.~\cite{bu,bm1}.
In order to reach higher precision to account for virtual leptons in
semi-leptonic decays, the theory has to be further extended and the
corresponding effective lagrangian evaluated.  This was achieved in
ref.~\cite{knrt}, and the decays $\pi_{l2}$ and $K_{l2}$ studied in
great detail.  Another interesting process that has been considered is
the beta decay of the charged pion~\cite{cknp}, which would lead to 
the determination of the Kobayashi-Maskawa matrix element $V_{ud}$ at
an unpredented level of accuracy, provided sufficiently accurate experimental
information is available.

The presence of the anomaly in QCD is accounted for in the
effective theory by the inclusion of the Wess-Zumino-Witten term which
arises at $O(p^4)$.  By including electromagnetism as an external source,
this accounts for the observed neutral pion decay into two photons.
This is often referred to as the anomalous or the odd intrinsic parity
sector. Recent developments in this field include the evaluation of the 
generating functional in this sector at $O(p^6)$~\cite{efs}
and also the inclusion of virtual photons to yield the generating functional at 
$O(e^2 p^4)$~\cite{am}.  Implications to the neutral pion lifetime
have been considered in some detail~\cite{am,gbh} which will be measured
to high accuracy at Jefferson Laboratory at the experiment PrimEx.

\section{Baryon chiral perturbation theory}
Baryon chiral perturbation theory in the modern era was first formulated
in~\cite{gss}.  This was a relativistic formulation, which faced the
technical difficulty in that no chiral power counting was possible when
the baryon was included in this framework.  Inspired by developments in
the field of heavy-quark physics, a version called heavy baryon chiral
perturbation theory was proposed, a formalism within which it was possible
to restore chiral order~\cite{jm}.  
This suffered from the deficiency that it was
no longer manifestly Lorentz invariant.  This showed up in the analysis of
certain form factors where analyticity properties were destroyed.  
A recent variant which simultaneously accounts for manifest Lorentz invariance
and chiral order has been developed~\cite{bl}. 
For related work that preceded this, see~\cite{et}.  
The main feature that is addressed is that in the chiral limit, the
baryon still remains massive.  

This infra-red regularized baryon chiral perturbation theory
is formally very interesting, and is also the basis of computations of
observables and much work is being carried out in comparing the consistency
of this framework and experimental information.  For instance the pion-nucleon
scattering amplitudes have been studied in~\cite{bl2,te}.
The corner stones of these studies include the corrections to
the Goldberger-Treiman relation and estimates of the sigma term.
Predictions for the masses of the
baryon octet in heavy baryon chiral perturbation theory were presented
in~\cite{bkmzeit} and a comparison between the heavy baryon 
and infra-red versions of the theory may be found in~\cite{et2}.  
Magnetic moments have been considered in the heavy baryon case in 
ref.~\cite{ms}, and work is in progress for the infra-red case~\cite{ellis}.
For results on form factors, see~\cite{kume}.
Despite many successes, a coherent picture of the nature of convergence
of baryon chiral perturbation theory, especially in the full $SU(3)$ is
yet to emerge.  Difficulties associated with iso-spin breaking in pion-nucleon
scattering need to be dealt with in a systematic manner, especially when
fresh data from pion photo-production is made available.

\section{Additional topics}

It has been pointed out earlier that the chiral predictions for pion
scattering lengths is a sensitive probe of the ground state of QCD and
chiral symmetry breaking therein.  Furthermore, the properties of the
ground state and the non-perturbative sector of QCD has important implications
for the chiral expansion, and in particular for the values of some of
the low-energy constants, and for those process which involve the axial
anomaly.  Indeed, it follows that the study of the global properties of
the QCD Lagrangian alone will not suffice for pinning down the chiral
expansion precisely.  One of the additional topics we consider is the
$N_c^{-1}$ expansion.  We also briefly mention the topic of simulations
on the lattice that might allow one to derive estimates for some low-energy
constants.  Finally, we comment on the thermodynamics of QCD and the
limits in which chiral symmetry is able to essentially determine its
structure.

\subsection{Large $N_c$}
An extremely interesting and subtle aspect of chiral perturbation theory
arises when one is trying to address the issue of the $N_c^{-1}$ expansion.
In particular, the role of the $\eta'$ is modified in this approach, and
one must understand the dynamics behind the Okubo-Zweig-Izuka rule.
Right from the original work in modern chiral perturbation theory, this
has been highlighted.  Recent work in applications of the expansion 
may be found in, e.g., ref.~\cite{bm2,kl}.  Indeed, we had pointed out
earlier that one of the low-energy constants $L_4$ which is suppressed
in the $N_c^{-1}$ expansion has now been estimated from $\pi K$ sum rules,
and supports earlier estimates from the Zweig rule.  

\subsection{Chiral perturbation theory and the lattice}
A relatively new area of research has been the one in which 
one may try to evaluate the low-energy constants using lattice simulations.
For instance, one may wish to evaluate the ratio of decay constants
$F_K/F_\pi$ on the lattice which would give an estimate for one of
the low-energy constants.  Furthermore, in chiral perturbation theory
one may study the quark mass dependence of physical observables at a given
order, which can be done independently on the lattice.  However, problems
remain in trying to reach small quark masses on the lattice and one would
have to wait some more years to have a reliable estimate of the 
low-energy constants using this approach.  There is also more recent
work in the field which is not discussed here.   However, it may be noted
that the lattice simulations might be able to test theoretical
ideas regarding the large $N_c$ expansion in a novel manner.  Of special
interest would be the possibility of testing the mixed expansion proposed
in ref.~\cite{bm2}.  Note also that there
has also been an effort to determine the $I=2$ pion scattering length
$a^2_0$ on the lattice~\cite{aoki},
for which we now have precise chiral predictions and measurements
from E-865.

\subsection{Chiral perturbation theory and thermodynamics}
Another rich field is that of the behaviour of chiral
perturbation at finite temperature.  The foundations of
the subject were laid in~\cite{glthermo}.  
There it has been shown that for small quark masses, low temperatures
and large volumes, the strong interaction partition function is
essentially fixed by chiral symmetry.  In addition, the law governing
the temperature dependence of the quark condensate has been determined.
Its dependence on the quark mass has also been computed.
The formal similarities between the finite temperature and the lattice
simulations provide useful checks for the latter.             
The nature of pions at finite temperature~\cite{toublan},
in particular the expansions for masses and decay constants have
also been studied.
Of related interest is the $\pi^0\to 2\gamma$ rate at finite 
temperature~\cite{ptt}.  

\section{Outlook}

In this talk I have outlined what I consider to be some specially interesting
topics.  The list of references provided in this written version of the talk
is not a complete list by any means in this
very active field of research, but a list of representative papers that
might give the reader a flavor of the subject.  What is specially striking is
the fact that effective field theories can be as predictive as they have been
in this context.  The criterion of renormalizability which is associated with
predictivity cannot be applied to this field.  I have also tried to bring to
the attention of the reader to areas where there is much scope for active
research in the near future.  I have not discussed some important topics,
e.g.,  non-leptonic weak decays,
the decay $\eta\to3\pi$~\cite{al}, etc., which represent important challenges.
The $NN$ system being significantly more involved has not been addressed
in this talk, although there is considerable amount of work in the field.

In addition to computing more interesting processes to two-loop accuracy in
chiral perturbation theory, and accounting for electro-magnetic corrections,
it might also be useful to look into techniques based on axiomatic field
theory and disperion relations which are completely general and are 
independent of the underlying dynamics and of effective field theory.
These techniques could yield additional constraints and provide consistency
checks both on the effective lagrangian computations, and on the analysis
of experimental information.  The successful application of these methods
has already been demonstrated in the pion-pion and pion-Kaon scattering
systems, and could definitely be extended to other settings.
There has also been a call for improving the experimental picture in
several sectors that would test chiral predictions, including photon
induced reactions, and those in which Kaons are involved in a significant 
manner~\cite{jgtalk}.

It might also be worth the while to revisit several arguments that 
are available in the literature for the inevitability of spontaneous
symmetry breaking of the axial-vector symmetries of QCD, including
results due to Vafa-Witten, Banks-Casher (reviewed in ref.~\cite{ge1})
and to see if any fresh insights can be found.
Finally, we note that
after this talk was given, a series of comprehensive discussions
has appeared on the archives~\cite{hl3papers}.
\acknowledgements{I thank the Department of Science and Technology, 
Government of India  for support
under the project entitled ``Some aspects of low-energy 
hadronic physics,''  the organizers of QCD2002, 
with special thanks to Pankaj Jain 
for inviting me to present this review,
and B. Moussallam for reading the manuscript.}

\end{document}